\documentclass[12pt,a4paper]{article}
\usepackage{amsfonts,latexsym}
\usepackage{graphicx,color}
\usepackage{dcolumn}
\usepackage{graphicx}
\usepackage{amsmath}
\usepackage{amsfonts}
\usepackage{amssymb}

\renewcommand{\thefootnote}{\fnsymbol{footnote}}

\begin{document}

\vspace{12mm}

\begin{center}
{{{\Large {\bf Quasibound states of massive scalar \\around the Kerr black hole}}}}\\[10mm]

Yun Soo Myung\footnote{e-mail address: ysmyung@inje.ac.kr}\\[8mm]

{Institute of Basic Sciences and Department  of Computer Simulation,  Inje University, Gimhae 50834, Korea\\[0pt]}

{and Asia Pacific Center for Theoretical Physics, Pohang 37673, Korea}

\end{center}
\vspace{2mm}

\begin{abstract}

\end{abstract}
\vspace{5mm}
 In this work, we obtain  quasibound states of a massive scalar around the Kerr black hole.
 A massive scalar propagating around the Kerr black hole induces   the far-region potential with either a trapping well or a tiny well, which produces a peak of resonance or
a bound state in the far-region. The former represents superradiant instability, while the latter could accommodate superradiant instability via a positive $\omega_I$.  If the scalar's  energy is really quantized around a tiny well,  an ultralight boson could be expressed in terms of  a hydrogen-like bound state, leading to  a gravitational atom when incorporating superradiant instability. If there are no wells outside the outer horizon, their far-region wave functions  represent purely  bound state without node. Finally, we discriminate between superradiant instability and stability.

\vspace{1.5cm}

\hspace{11.5cm}
\newpage
\renewcommand{\thefootnote}{\arabic{footnote}}
\setcounter{footnote}{0}


\section{Introduction}
Direct detection of the gravitational waves (GWs) sourced from several binary coalescences~\cite{LIGOScientific:2016aoc,LIGOScientific:2021sio} has shed light on a new era of the astronomy.  Superradiance refers to the possibility of energy and
angular momentum extraction from the rotating black hole when the superradiance condition ($\omega <\omega_c$) is
satisfied~\cite{Press:1972zz,Damour:1976kh,Brito:2015oca}.  If a massive scalar is impinging upon a rotating black hole, its mass $\mu$ acts as a reflector and leads to superradiant instability~\cite{Cardoso:2004nk}.
Interestingly,  ultralight bosons~~\cite{Arvanitaki:2009fg,Arvanitaki:2010sy} being dark matter candidates may  trigger superradiant instability of rotating black holes and form bosonic clouds to emit GWs. These GWs could be detected with future gravitational wave observations.

It was proposed that the superradiant instability of an ultralight boson could produce a gravitational atom like a hydrogen atom which  consists of  the Kerr black hole (mass $M$ and  rotation parameter $a$) and a surrounding boson cloud in quantum states~\cite{Kodama:2011zc}.
In the far-region of $r$, the radial part of the Klein-Gordon equation for the ultralight boson with small mass ($M\mu\sim M\omega \ll1$)  becomes the Schr\"{o}dinger-type equation for the electron in a hydrogen atom and it could be solved exactly in terms of the confluent hypergeometric function $U(\tilde{p},s;x)$.
In this direction, Detweiler has obtained the maximum growth rate [$\omega_{\rm I}\sim a(M\mu)^9/(24M^2)$] for superradiant instability  by making use of two asymptotic wave functions~\cite{Detweiler:1980uk}:
 near-horizon wave function given by the hypergeometric function ${}_2F_1(\tilde{a},\tilde{b},\tilde{c};x)$ and  far-region wave function expressed by $e^{-x/2}x^{l+1}U(l+1-\nu,2l+2;x)$. Here, the replacement of $l+1-\nu \to -n-\delta \nu$ was done  by considering a small shift from the energy level of the hydrogen-like wave function.  In this case. one expects to find a free oscillation around a bound state by insisting that
$n$ is  a non-negative integer and a small complex number $\delta\nu(|\delta\nu|\ll1)$ represents a deviation from the hydrogen-like wave function. The latter  describes a slowly growing superradiant instability. This replacement was inspired  by a proposal that $ \omega_{\rm I}$  induces a complex $\delta\nu$.

In all of superradiance studies, the computation of bosonic bound states is a main part.
It is important  to explore a connection  between explicit form of bound states and far-region potential.
In this work, we wish to construct  quasibound states of  a massive scalar around the Kerr black hole explicitly.
Even though the  equation is the same as  that  for the hydrogen atom, its energy is positive and continuous, compared to the negative quantized energy for the hydrogen atom.
 A massive scalar with mass $\mu$ propagating around the Kerr black hole induces  usually the far-region potentials with  a trapping well ($\mu\le l$), a  tiny well ($\mu \ll l$), and no well $(\mu>l$)~\cite{Myung:2022biw,Myung:2022yuo}. In general, the first two produce a peak (resonance) or
a bound state in the far-region only when imposing the asymptotic bound state (abs) condition of $\omega<\mu$.
The other wave function in the overlapping region  will be bounded or unbounded.
If the scalar's  energy $\omega^2$ is  quantized around any well especially, its wave function  describes a hydrogen-like bound state.
For  an ultralight boson, it may  produce a gravitational atom due to superradiant instability.
In case of  $\mu>l$,  there is no well outside the outer horizon  and  thus, their far-region wave functions all represent purely  bound state without node.
Furthermore,  we distinguish  between superradiant instability and stability.

\section{Massive scalar around  Kerr black holes }
A scalar perturbation $\Phi$ with mass $\mu$ obeys the Klein-Gordon equation~\cite{Brito:2015oca}
\begin{equation}
(\bar{\nabla}^2-\mu^2)\Phi=0\label{phi-eq1}
\end{equation}
on the background of Kerr black holes expressed in term of  Boyer-Lindquist coordinates
\begin{eqnarray}
ds^2_{\rm K} &\equiv& \bar{g}_{\mu\nu}dx^\mu dx^\nu \nonumber \\
&=&-\frac{\Delta}{\rho^2}\Big(dt -a \sin^2\theta d\phi\Big)^2 +\frac{\rho^2}{\Delta} dr^2+
\rho^2d\theta^2 +\frac{\sin^2\theta}{\rho^2}\Big[(r^2+a^2)d\phi -adt\Big]^2 \label{KBH}
\end{eqnarray}
with
\begin{eqnarray}
\Delta=r^2-2Mr+a^2,~ \rho^2=r^2+a^2 \cos^2\theta,~{\rm and}~a=\frac{J}{M}.
 \label{mps}
\end{eqnarray}
Here, $M$ and  $J$ denote  the mass and angular momentum of the Kerr black hole with units of $G=c=\hbar=1$.
The outer and inner horizons are obtained   by requiring  $\Delta=(r-r_+)(r-r_-)=0$ as
\begin{equation}
r_{\pm}=M\pm \sqrt{M^2-a^2}.
\end{equation}
We decompose the scalar perturbation
as
\begin{equation}
\Phi(t,r,\theta,\phi)=e^{-i\omega t+im\phi} S_{lm}(\theta)
 R_{lm}(r), \label{sep}
\end{equation}
where   $S_{lm}(\theta)$ denotes the spheroidal harmonics with $l$ spheroidal harmonic index and $m$ azimuthal harmonic index
and $R_{l m}(r)$ represents the radial wave function.

Plugging (\ref{sep}) into (\ref{phi-eq1}), we have
an  angular  equation
\begin{eqnarray}
&& \frac{1}{\sin \theta}\frac{d}{d\theta}\Big(
\sin \theta
\frac{d}{d\theta} S_{lm}(\theta) \Big )+ \left [K_{lm}+a^2(\mu^2-\omega^2)\cos^2\theta-\frac{m^2}{\sin ^2{\theta}} \right ]S_{lm}(\theta) =0
\label{wave-ang1}
\end{eqnarray}
with  separation constant $K_{lm}=l(l+1)+\sum_{k=1}^{\infty}c_ka^{2k}(\mu^2-\omega^2)^k$.
The radial Teukolsky  equation is given by
\begin{eqnarray}
\Delta \frac{d}{dr} \Big( \Delta   \frac{d}{dr}R_{l m}(r) \Big)+U(r)R_{l m}(r)=0
\label{wave-rad}
\end{eqnarray}
with the potential
\begin{eqnarray}
U(r)=[\omega (r^2+a^2)-am]^2+\Delta[2am\omega-\mu^2(r^2+a^2)+a^2(\mu^2-\omega^2) -K_{lm}].\label{U-pot}
\end{eqnarray}
Let us introduce the tortoise coordinate $r_*$
\begin{equation}
r_*=\int\frac{(r^2+a^2)dr}{\Delta}=r+\frac{2Mr_+}{r_+-r_-}\ln\Big(\frac{r}{r_+}-1\Big)-\frac{2Mr_-}{r_+-r_-}\ln\Big(\frac{r}{r_-}-1\Big).
\end{equation}
In this case, an interesting region of $r\in[r_+,\infty)$ could be mapped into the whole region of $r_*\in(-\infty,\infty)$. This implies that
the inner region of $r<r_+$ is not our concern for studying superradiant instability. Also, we have $r_*\simeq r$ for $r\gg r_+$ and $r_*\simeq r_++2Mr_+/(r_+-r_-)\ln(r/r_+-1)$ for $r\sim r_+$.
The Teukolsky equation (\ref{wave-rad}) leads to
the Schr\"odinger-type equation with $\Psi_{l m}=\sqrt{r^2+a^2} R_{l m}$ and the energy $\omega^2$ as
\begin{equation}
\frac{d^2\Psi_{l m}(r_*)}{dr_*^2}+[\omega^2-V_{\rm K}(r)]\Psi_{l m}(r_*)=0. \label{sch-eq}
\end{equation}
Here, the scalar potential $V_{\rm K}(r)$ takes a complicated form~\cite{Zouros:1979iw,Arvanitaki:2010sy}
\begin{eqnarray}
V_{\rm K}(r)&=&\omega^2\nonumber \\
&+&\frac{\Delta \mu^2}{a^2+r^2}-\Big(\omega-\frac{am}{a^2+r^2}\Big)^2
-\frac{\Delta}{(a^2+r^2)^2}\Big[2am\omega -K_{lm}+a^2(\mu^2-\omega^2)\Big]\label{c-pot1} \\
&-&\frac{3\Delta^2r^2}{(a^2+r^2)^4}+\frac{\Delta[\Delta+2r(r-M)]}{(a^2+r^2)^3}, \nonumber
\nonumber
\end{eqnarray}
where the second  line represents $-U(r)/(r^2+a^2)^2$, whereas the third line denotes the effect of introducing the coordinate $r_*$.
We stress that the former governs the near-horizon  and asymptotic limits since $V_{\rm K}(r\to r_+) = \omega^2-(\omega-\omega_c)^2$ with the critical frequency $\omega_c=m a/(r_+^2+a^2)$ and $\Delta(r\to r_+)= 0$, while $V_{\rm K}(r\to \infty)=\mu^2$.

At this stage, we would like to mention two conditions for superradiant instability: superradaince condition $\omega<\omega_c$ and abs condition $\omega<\mu$.
It is clear  that if a massive scalar with mass $\mu$ is scattered off by a rotating black hole with $M$ and $a$, then for $\omega<\mu$, the superradiance with $\omega<\omega_c$ has unstable modes because the scalar mass term works effectively as a reflecting mirror. Importantly, the other condition for superradiant instability is given by existing  a local minimum ($\bullet$, a positive well)~\cite{Zouros:1979iw}.
In  case of Kerr black holes, this  is given by either a trapping well located outside the outer horizon or a tiny well located far from the outer horizon~\cite{Myung:2022biw,Myung:2022yuo}.
For small mass $M\mu\ll1$ and $M\omega \ll1$, one obtains $K_{lm}\simeq l(l+1)$ by neglecting $\sum_{k=1}^{\infty}c_ka^{2k}(\mu^2-\omega^2)^k$. In the far-region ($r\gg r_+$) where one takes $r_*\simeq r$,   we obtain an equation to describe  $\Psi_{l}(=r R_{l})$ from (\ref{sch-eq})  as
\begin{equation}
\Big[\frac{d^2}{dr^2}+\omega^2-V_{\rm far}(r)\Big]\Psi_{l}(r)=0 \label{far-eq}
\end{equation}
with the far-region potential
\begin{equation}
V_{\rm far}(r)=\mu^2-\frac{2M\mu^2}{r}+\frac{l(l+1)}{r^2}=\mu^2\Big[1-\frac{2M}{r}+\frac{l(l+1)}{\mu^2r^2}\Big].
\end{equation}
We note that $V_{\rm far}(r)$ is obtained for $K_{lm}$ when expanding $V_{\rm K}(r)$ for large $r$ and keeping it up to $(1/r)^2$-order.
Thus, $V_{\rm far}(r)$ is just the far-region potential when choosing $K_{lm}=l(l+1)$.
As we mentioned before, the other condition for superradiant instability is given by the presence of a local minimum ($\bullet$) at $r= r_0>r_+$.
From the condition of $V'_{\rm far}(r)|_{r=r_0}=0$, we find a local minimum at $r=r_0$ where
\begin{equation}
r_0=\frac{l(l+1)}{\mu^2}
\end{equation}
with $M=1$.
For Kerr black holes, this  is described  by either a trapping well ($\mu \le l$) located outside the outer horizon or a tiny well ($\mu \ll l$) located far from the outer horizon. For $r_0<r_+(\mu>l)$, any well does not appear in the relevant region of $r\in [r_+,\infty)$.

Introducing the definitions as
\begin{equation}
 x=2kr,\quad k=\sqrt{\mu^2-\omega^2},\quad \nu=\frac{M\mu^2}{k}, \label{l-def1}
\end{equation}
Eq.(\ref{far-eq}) becomes the Whittaker's equation
\begin{equation}
\frac{d^2\Psi_{l}(x)}{dx^2} +\Big[-\frac{1}{4}+\frac{\nu}{x}-\frac{l(l+1)}{x^2}\Big]\Psi_{l}(x)=0, \label{c-eq}
\end{equation}
which is the same equation as Eq.(\ref{h-eq})  for an electron in the hydrogen atom.

Considering $\Psi_{l}(x)=e^{-x/2} x^{l+1} v(x)$, Eq.(\ref{c-eq}) leads to the Kummer's equation for $v(x)$ as
\begin{equation}
x\frac{d^2v(x)}{dx^2}+[(2l+1)+1-x]\frac{dv(x)}{dx}+[\nu-(l+1)]v(x)=0 \label{kumm1}
\end{equation}
whose solution is given exactly by the (second kind) confluent  hypergeometric function  as
\begin{eqnarray}
v_U(x)=U(l+1-\nu,2l+2;x) \label{wavef-1},\quad x=2kr.
\end{eqnarray}
Its two asymptotic forms for  $x\gg1 $and $x\ll1$ with $(2l+2) \not\in \mathbb{Z}$ are given by
\begin{eqnarray}
&&U(l+1-\nu,2l+2;x\gg1)\to x^{\nu-l-1}\Big[1-\frac{(\nu-l-1)(l+\nu)}{x}+{\cal O}\Big(\frac{1}{x}\Big)^2\Big], \label{uexp1} \\
&&U(l+1-\nu,2l+2;x\ll1)\to \frac{\Gamma(-2l-1)}{\Gamma(-l-\nu)}+\frac{\Gamma(2l+1)}{\Gamma(l+1-\nu)}x^{-(2l+1)}+\cdots. \label{uexp2}
\end{eqnarray}
Here, for $(2l+2) \in \mathbb{Z}$ (our case), the relation of Eq.(\ref{uexp2}) is no longer true because  $\Gamma(-2l-1)$ blows up.
Eq.(\ref{uexp1})  is used to derive an asymptotic wave function as
\begin{equation}
\Psi^{\rm A}_l(x) \sim e^{-x/2}x^\nu,
\end{equation}
which always shows asymptotic  bound state for $\omega<\mu$.  On the other hand, Eq.(\ref{uexp2}) could be employed to match  the near-horizon wave function at the overlapping (ol) region of $kM\ll x \ll 1$ for the ultralight boson only. In case of $(2l+2) \not\in \mathbb{Z}$, the wave function $R^{\rm ol}_l(x)=\Psi^{\rm ol}_l(x)/x$ may take the form~\cite{Detweiler:1980uk}
\begin{equation}
R^{\rm ol}_l(x) \sim \frac{\Gamma(-2l-1)}{\Gamma(-l-\nu)}x^{l}+\frac{\Gamma(2l+1)}{\Gamma(l+1-\nu)}x^{-l-1}+\cdots.\label{psie-1}
\end{equation}

However, the (first kind) confluent  hypergeometric function $v_{F}(x)={}_1F_1(l+1-\nu,2l+2;x)$ is not a proper solution.
Here, we denote two far-region wave solutions:
\begin{equation}
\Psi^\nu_{lU}(x)=e^{-x/2}x^{l+1}v_U(x),\quad \Psi^\nu_{lF}(x)=e^{-x/2}x^{l+1}v_{F}(x). \label{t-wavef}
\end{equation}
At this stage, it is important to know that  there is no restriction on $\nu$ after  imposing the abs condition $\omega<\mu$ for superradiant instability.
In general, it would be  noted that  $\nu$ is not an integer but an irrational number.  This contrasts sharply  to the corresponding $\rho_0=n(n\in \mathbb{N})$ in Eq.(\ref{kumm2}) for the hydrogen atom.
That is, it is special to take $l+1-\nu=-\tilde{n}(\tilde{n}\in \mathbb{N})$ in the massive scalar propagation around the Kerr black hole. Considering Eq.(\ref{uexp1}), $U(l+1-\nu,2l+2;x)$ is an increasing (decreasing) function for $\nu> l+1(\nu<l+1)$ approximately. The case of $\nu=l+1(\tilde{n}=0)$ may represent a boundary between increasing  and decreasing functions, giving $U(0,2l+2;x)=1$.  Therefore, a condition for the superradiant instability is given by $\nu>l+1$, while the condition for superradiant stability is proposed as $\nu<l+1$~\cite{Myung:2022biw}. It is worth noting that $\Psi^\nu_{lU}(x)$ describes either a peak (resonance) or a bound state when requiring the abs condition of $\omega<\mu$.

Imposing  $\nu=\hat{n}~(\hat{n}\in \mathbb{N}$, principle quantum number), one may  find that the energy  $\omega^2$  is quantized as~\cite{Detweiler:1980uk}
\begin{equation}
\omega^2=\mu^2-\frac{(M\mu^2)^2}{\hat{n}^2}\to \omega_{\hat{n}} \simeq \mu\Big(1-\frac{M^2\mu^2}{2\hat{n}^2}\Big), \label{omega-R}
\end{equation}
where $\omega_{\hat{n}}$ denotes a spectrum of hydrogen-like bound states around a positive well.
For $M\mu \ll \hat{n}$, one has the relation of $\omega_{\hat{n}}\sim \mu \ll \hat{n}/M$.
This implies that one  is able  to obtain the condition ($M\mu\sim M\omega_{\hat{n}} \ll1$) for an  ultralight boson  when selecting $\nu=\hat{n}$.
For $l+1-\hat{n}=-\tilde{n}$ , $\Psi^{\hat{n}}_{ lU}(x)$ and $\Psi^{\hat{n}}_{ lF}(x)$  in (\ref{t-wavef}) may describe the hydrogen-like bound state.
This is equivalent to achieving two bound states in overlapping and asymptotic regions.
Detweiler has found the maximum growth rate ($\omega_{\rm I}$) for superradiant instability of an ultralight boson with $l=m=1$ around the Kerr black hole  with $a/M\sim 1$ by introducing analytic functions in two asymptotic regions~\cite{Detweiler:1980uk}. In this case, its real part $\omega_{\rm R}$ is given by $\omega_{\hat{n}}$ in Eq.(\ref{omega-R}).

It is important to  to distinguish between   scalar wave function around Kerr black holes and hydrogenic wave function.
For this purpose, it is interesting to introduce the radial equation for an electron $(E<0)$ bound  to nucleus with $u_l(r)=rR_l(r)$ and $r\in[0,\infty]$ as
\begin{equation}
-\frac{\hbar^2}{2m}\frac{d^2u_l(r)}{dr^2}+V_{\rm h}(r) u_l(r)=Eu_l(r), \label{hydrogen}
\end{equation}
where the hydrogen potential is given by
\begin{equation}
V_{\rm h}(r)=-\frac{e^2}{4\pi \epsilon_0}\frac{1}{r}+\frac{\hbar^2}{2m}\frac{l(l+1)}{r^2}=\frac{e^2}{4\pi \epsilon_0a_0}\Big[-\frac{a_0}{r}+\frac{l(l+1)}{2}\frac{a_0^2}{r^2}\Big]
\end{equation}
with $a_0$ the Bohr radius. The potential energy of the electron is given by  $-e^2/(4\pi\epsilon_0 a_0)=-4.347 \times 10^{-18}$[J].
Eq.(\ref{hydrogen}) could be rewritten as the Whittaker's equation
\begin{equation}
\frac{d^2u_{l}(\rho)}{d\rho^2} +\Big[-\frac{1}{4}+\frac{\rho_0}{\rho}-\frac{l(l+1)}{\rho^2}\Big]u_{l}(\rho)=0, \label{h-eq}
\end{equation}
where
\begin{equation}
\rho=\sqrt{\frac{-8m E}{\hbar^2}} r, \quad \rho_0=\frac{e^2}{4\pi \epsilon_0 \hbar}\sqrt{\frac{-m}{2E}}.
\end{equation}
Introducing  $u_{l}(\rho)=e^{-\rho/2} \rho^{l+1} z(\rho)$ after imposing the inner boundary condition that the electron wave function must be regular at the origin, Eq.(\ref{h-eq}) leads to the Kummer's equation for $z(\rho)$ as
\begin{equation}
\rho\frac{d^2z(\rho)}{d\rho^2}+[(2l+1)+1-\rho]\frac{dz(\rho)}{d\rho}+[\rho_0-(l+1)]z(\rho)=0.\label{kumm2}
\end{equation}
Choosing $z(\rho)=\sum_{j=0}^\infty c_j\rho^j$ and imposing the termination of series  at $c_{j_{\rm max}}=0$ to avoid a solution like  $z(\rho)\sim e^{\rho}$,
one finds
\begin{equation}
\rho_0=n,\quad E_n=-\Big[\frac{m}{2\hbar^2}\Big(\frac{e^2}{4\pi \epsilon_0}\Big)^2\Big]\frac{1}{n^2}=\frac{E_1}{n^2},\quad \rho=\frac{2}{a_0 n}r,\label{quant-C}
\end{equation}
where $n=1,2,3,\cdots$ is the principal quantum number, $E_1=-2.18\times 10^{-18}$[J] denotes the energy for the ground state, and $a_0=4\pi\epsilon_0\hbar^2/me^2=0.58\times 10^{-10}$[m]  is the Bohr radius. We note that the  termination condition for series is equivalent to imposing the abs condition of $\omega<\mu$.

Then, an appropriate  solution to Eq.(\ref{kumm2}) is given  by the Laguerre polynomial $L(a,b;\rho)$ as
\begin{eqnarray}
z_L(\rho)&=& L(n-l-1,2l+1;\rho) \label{wavef-1}
\end{eqnarray}
together with the existence condition of $n\ge l+1$. This means that the other case of $n<l+1$ is not allowed for a hydrogenic bound state and $n=l+1$ is the bottom line. Also, we note that $z_U(\rho)=U(-n+l+1,2l+2;\rho)$ and $z_F(\rho)={}_1F_1(-n+l+1,2l+2;\rho)$ become  proper solutions for the electron.
So, we obtain  three proper wave functions for a hydrogen atom as
\begin{equation}
u^n_{lL}=e^{-\rho/2}\rho^{l+1}z_L(\rho),\quad u^n_{lU}=e^{-\rho/2}\rho^{l+1}z_U(\rho),\quad u^n_{lF}=e^{-\rho/2}\rho^{l+1}z_F(\rho),
\end{equation}
where all these describe hydrogenic bound states for $n\ge l+1$. Here, one denotes $l=0,1,2,3,\cdots$ by {\it s, p, d, f}, $\cdots$ states. Also, $n-l-1$ represents number of nodes.
\begin{figure*}[t!]
   \centering
  \includegraphics{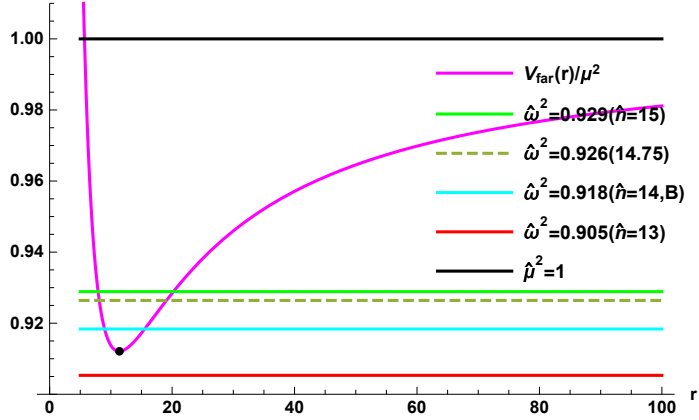}
\caption{ Positive potential $V_{\rm far}(r)/\mu^2$ with $M=1,\omega=3.85,\mu(=4<l),l=13,\nu=14.75,k=1.09$ represents a  peak (resonance) because of energy $\omega^2(=14.82)>0$. A local minimum ($\bullet$) is located at $(r_0=11.4,0.91)$. Here, we introduce $r\in[5,100]$, $\hat{\omega}^2=\omega^2/\mu^2=0.926$, and $\hat{\mu}^2=\mu^2/\mu^2=1$. Other energies $\hat{\omega}^2$ are given by the energy quantization Eq.(\ref{omega-R}). }
\end{figure*}

\section{Quasibound states}
\subsection{Potential with a trapping well}
Even though their equations (\ref{c-eq})[(\ref{kumm1})]  and (\ref{h-eq})[(\ref{kumm2})] are the same, their potentials and energies are different.
Let us consider a potential displayed in Fig. 1. This  potential of $V_{\rm far}(r)$ has a trapping well ($\bullet$, local minimum located at $r=11.4$), but it could not describe the ultralight boson with $M\mu \ll1$ and $M\omega \ll1$~\cite{Zouros:1979iw}. This potential is purely repulsive (i.e., $V_{\rm far}(r)>0$), so there are no hydrogenic bound states. Like a bound state, a quasibound state is strongly concentrated in the trapped region of space, but it lies within the continuous energy for the free states. Therefore, the quasibound state is a state between bound and free states.   In a continuum  state of $V_{\rm far}(r_0)<\omega^2(=14.82)<\mu^2$, the reduced energy $\hat{\omega}^2(=0.926)$ is between $\hat{\omega}^2_{\hat{n}=14}$ and $\hat{\omega}^2_{\hat{n}=15}$.  Also,  $\nu$ and $k$ have continuous values.
One regular solution [$v_U(r),~r\in[5,100]$] to Eq.(\ref{kumm1}) is given by the (second-kind) confluent  hypergeometric function $U^{14.75}(\tilde{p}=-0.75,28;2.17r)$ which is an increasing function [(Left) Fig. 2] because of $\tilde{p}<0$, while the other solution [$v_F(r),~r\in[11,100]$] is  the (first-kind) confluent  hypergeometric function ${}_1F_1(-0.75,28;2.17r)$ which negatively blows up around $r=80$. This implies that $v_F(r)$ is not a proper solution.
The wave function [$\Psi^{14.75}_{13U}(r)=5.14\times 10^4 e^{-1.085 r} r^{14} v_U(r)$]  describes  a peak (resonance) as shown in (Right) Fig. 2~\cite{Myung:2022biw,Myung:2022yuo}. This peak with height $5\times 10^{14}$  is regarded as  a resonance. The incident waves propagate to  the black hole and spends a long time trapped  in the positive well to make a peak, before they  are eventually leaking out and escaping  to infinity. This peak is due to the increasing function of $v_U(r)$.
\begin{figure*}[t!]
   \centering
  \includegraphics{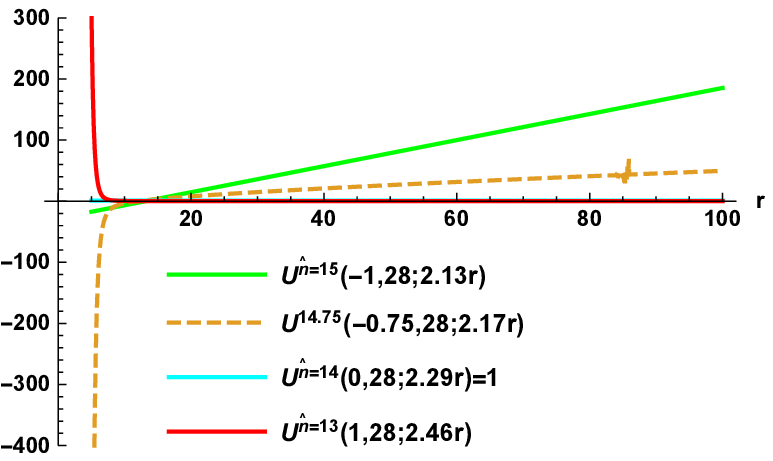}
   \hfill%
  \includegraphics{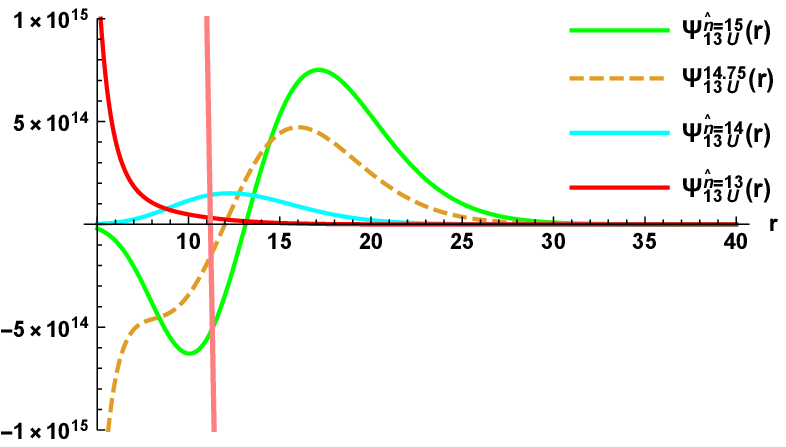}
\caption{(Left) Confluent hypergeometric functions $v_U(r)=U^{14.75}(-0.745,28;2.17r)$ and $U^{\hat{n}=15}(-1,28;2.13r)$ are increasing functions of $r\in[5,100]$, while $U^{\hat{n}=14}(0,28;2.29r)=1$ and  $U^{\hat{n}=13}(1,28;2.46r)$ is a decreasing function for Fig. 1 with a trapping well. (Right)  Quasibound states  [$\Psi^{\hat{n}=15}_{13U}(r)$ and $\Psi^{14.75}_{13U}(r)$],  a half of a peak [boundary bound state, $\Psi^{\hat{n}=14}_{13U}(r)$], and a purely bound state [$\Psi^{\hat{n}=13}_{13U}(r)$] as  functions of $r\in[11.4$(pink line), 40]. $\Psi^{14.75}_{13U}(r)$ and $\Psi_{13U}^{\hat{n}=13}(r)$  are unbound wave functions, because they blow up as $r$ approaches 5 (left boundary). If $\Psi^{\hat{n}=15}_{13U}(r)$ is extended to the left boundary, it looks like a hydrogen-like bound state.  }
\end{figure*}
On the contrary, $\Psi^{14.75}_{13F}(r)=5.14\times 10^4 e^{-1.085 r} r^{14} v_F(r)$  denotes an inappropriate state  because it blows up negatively around $r=95$.
For a complex $\omega=3.85+10^{-6}i~(|\omega_{\rm I}|\ll \omega_{\rm R},~\omega_{\rm R}<m\Omega_H)$ with the same other parameters as in  Fig. 1, one obtains  its far-region complex solution
 \begin{eqnarray}
 \Psi_{13U}^{c,14.75}(r)&=& (5.14\times 10^{4} - 2.35 i) e^{(-1.085+3.55\times 10^{-6} i)r}r^{14} \nonumber \\
 &\times& U\Big(\tilde{p}=-0.75-4.8\times 10^{-5}i,28;(2.17-7.09\times 10^{-6}i)r\Big), \label{c-wavef}
 \end{eqnarray}
 whose real part Re[$\Psi^{c,14.75}_{13U}$] takes the same form of a peak ($\Psi^{14.75}_{13U}$) as in (Right) Fig. 2.  Here, one may try to rewrite $\tilde{p}$ in $U(\tilde{p},s;cr)$ as
 \begin{equation}
 \tilde{p}=-n-\delta\nu
 \end{equation}
with $n=1$ and $\delta\nu=-0.25+4.8\times 10^{-5} i(|\delta\nu|\simeq0.25)$. The latter complex number may  be interpreted as  a deviation from the $\hat{n}=14$ (upper decimal part of $\nu=14.75)$ hydrogen-like energy level. However, this state of superradiant instability could not be approximately described by a small shift from  hydrogen-like  energy level because $|\delta\nu|=0.25$ is not a small shift.

Especially, we focus on  the energy quantization  Eq.(\ref{omega-R}) with $\nu=15(\tilde{n}=-1)$ with the same  potential [Fig. 1] which includes a trapping  well ($\bullet$). But, we have $\omega=3.85516(\hat{\omega}^2=0.929)$, corresponding to the $\hat{n}=15$ quantized state. It is shown that $v^{\hat{n}=15}_U(r)=U(-1,28;2.13r)$ is  an increasing function [(Left) Fig. 2].
Its wave function $\Psi^{\hat{n}=15}_{13U}(r)=4.04\times 10^4 e^{-1.067 r} r^{14} v^{\hat{n}=15}_U(r)$  describes  a peak (resonance) as shown in (Right) Fig. 2.
For a complex $\omega=3.85516+10^{-6}i~(|\omega_{\rm I}|\ll \omega_{\rm R},~\omega_{\rm R}<m\Omega_H)$ with the same other parameters as in (Left) Fig. 1, one obtains  its far-region complex solution
 \begin{eqnarray}
 \Psi^{c,\hat{n}=15}_{13U}(r)&=& (4.04\times 10^{4} - 1.92 i) e^{(-1.067+3.61\times 10^{-6} i)r}r^{14} \nonumber \\
 &\times& U\Big(\tilde{p}=-1-5.08\times 10^{-5}i,28;(2.13-7.23\times 10^{-6}i)r\Big), \label{c-wavef}
 \end{eqnarray}
 whose real part Re[$\Psi^{c,\hat{n}=15}_{13U}$] takes the similar form of a peak ($\Psi^{\hat{n}=15}_{13U}$) in (Right) Fig. 2.  Here, one may try to rewrite $\tilde{p}$ in $U(\tilde{p},s;cr)$ as
 \begin{equation}
 \tilde{p}=-n-\delta\nu
 \end{equation}
with $n=1$ and $\delta\nu=5.08\times 10^{-5} i(|\delta\nu|\ll1)$. The latter complex number may  be interpreted as  a deviation from the $\hat{n}=15$ hydrogen-like energy level.

We note that the energy quantization  Eq.(\ref{omega-R}) with $\nu=14(\tilde{n}=0)$ corresponds to the $\hat{n}=14$ quantized state. In this case, $v^{\hat{n}=14}_U=U(0,28;2.29r)=1$ is  a constant function between $U(\tilde{p}<0,s;x)$ and $U(\tilde{p}>0,s;x)$ [(Left) Fig. 2].
Its wave function [$\Psi^{\hat{n}=14}_{13U}(r)=1.06\times 10^5 e^{-1.14 r}r^{14}$]  describes  a half of a peak (boundary bound state) as shown in (Right) Fig. 2. Also, for $\hat{n}=13<l+1(\tilde{n}=1)$, we would like to mention that $v^{\hat{n}=13}_U(r)=U^{\hat{n}=13}_{13}(1,28;2.46r)$ is a rapidly decreasing function [(left) Fig. 2] and  the wave function [$\Psi^{\hat{n}=13}_{13U}(r)=2.99\times 10^5 e^{-1.23 r}r^{14}v^{\hat{n}=13}_U(r)$]  represents  a purely bound state as shown in (Right) Fig. 2,  but it is not a hydrogen-like bound state. In addition, we would like to mention that two wave functions $\Psi^{\hat{n}=13}_{13U}(r)$ and $\Psi^{14.75}_{13U}(r)$ blow up as $r$ approaches 5 (left boundary).

Lastly, considering  the condition for  superradiant instability ($\tilde{p}<0$) and stability ($\tilde{p}>0$) for a given potential with a trapping well [Fig. 1]~\cite{Myung:2022biw},
two [$\Psi^{\hat{n}=15}_{13U}(r)$ and $\Psi^{14.75}_{13U}(r)$] show superradiant instability with peak, while $\Psi^{\hat{n}=13}_{13U}(r)$ indicates  a purely bound state
(superradiant stability).
Furthermore, the case of $\Psi^{\hat{n}=14}_{13U}(r)$ represents a boundary between superradiant instability and stability [boundary bound state with with $U(0,28;2.29r)$]. This might be so because
the potential $V_{\rm K}(r)$ in Eq.(\ref{c-pot1}) with a trapping well is an energy ($\omega^2$)-dependent potential.

\subsection{Ultralight boson with a tiny well}
Now we have a position to consider the ultralight boson with a tiny well.
Let us introduce  an ultralight boson with $M\mu=0.01 \ll 1$ and $M\omega=9.9998\times 10^{-3} \ll1~(\hat{\omega}^2=0.9996)$ located below the $\hat{n}=2$ quantized level. This case has a potential with a tiny well ($\bullet$) located at $r=20000$  where is far from the outer horizon [(Left) Fig. 3].
One regular solution is given by the confluent  hypergeometric function $U^{1.58}(\tilde{p}=0.42,4;0.00013r)$ which is a rapidly decreasing function because of $\tilde{p}>0$ [(Right) Fig. 4].
Its wave function [$\Psi^{1.58}_{1U}(r)=1.6\times10^{-8} e^{-0.00006 r} r^2 U^{1.58}(0.42, 4;
  0.00013r)$] represents a purely bound state [(Right) Fig. 4].
For a complex $\omega=9.9998\times 10^{-3}+10^{-19}i~(\omega_{\rm I}=\frac{a}{24M^2}(M\mu)^9\sim 10^{-19}\ll \omega_{\rm R})$ with the same other parameters as in (Left) Fig. 3, one has its far-region solution
 \begin{eqnarray}
 \Psi_{1U}^{c,1.58}(r)&=& (1.6\times 10^{-8} - 8.0 \times 10^{-21} i) e^{(-0.00006 + 1.58\times 10^{-17} i)r}r^{2} \nonumber \\
 &\times& U\Big(\tilde{p}=0.42 - 3.95 \times 10^{-13}i,4;(0.00013 - 3.16\times 10^{-17} i)r\Big), \label{c-wavef}
 \end{eqnarray}
 whose real part Re[$\Psi^{c,1.58}_{1U}$] takes the same form of $\Psi^{1.58}_{1U}(r)$ as in (Right) Fig. 4.
 However, one may not rewrite $\tilde{p}$ in $U(\tilde{p},s;cr)$ as $\tilde{p}=-n-\delta\nu$ because `0.42' is not a negative integer.
This shows a failure of obtaining $\tilde{p}=-n-\delta\nu$ for this ultralight boson. The other ultralight boson with $M\mu=0.01$ and $M\omega=9.999991 \times 10^{-3}[\hat{\omega}^2=0.999982(2.36)]$ indicates that the confluent hypergeometric function $U^{2.36}(-0.36,4;0.000085r)$ is a rapidly increasing function [(Right) Fig. 4], while its wave function
$\Psi^{2.36}_{1U}(r)$ represents a boundary bound state [(Right) Fig. 4]. Also,  this case indicates  a failure of obtaining $\tilde{p}=-n-\delta\nu$ for the ultralight boson.
We note that two wave functions $\Psi^{1.58}_{1U}(r)$ and $\Psi^{2.36}_{1U}(r)$  blow up as $r$ approaches 100 (left boundary).
\begin{figure*}[t!]
   \centering
  \includegraphics{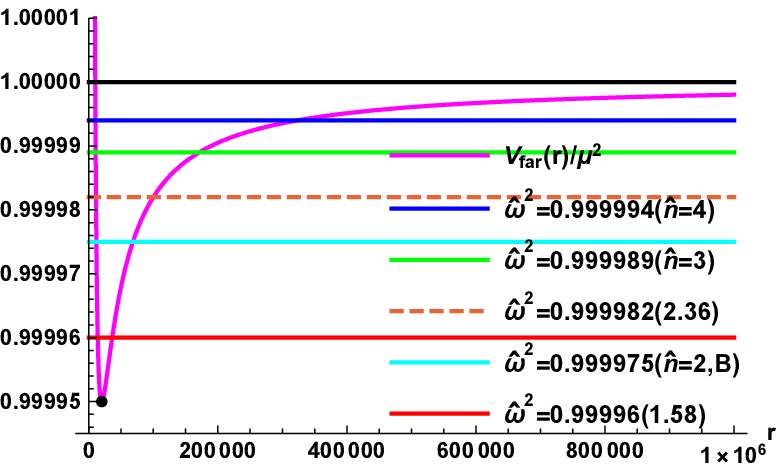}
  \hfill%
  \includegraphics{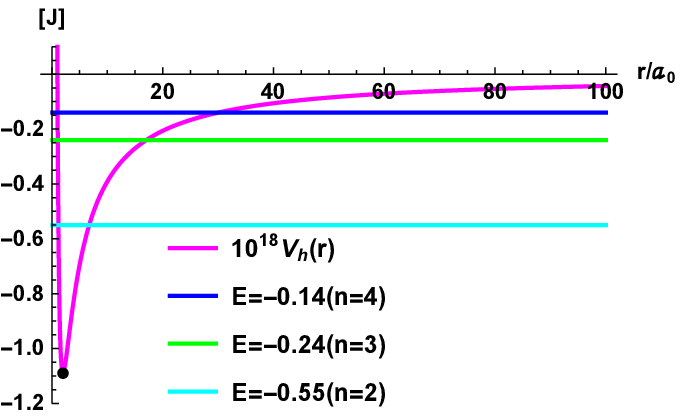}
\caption{(Left) Positive potential $V_{\rm far}(r)/\mu^2$ with $M=1,\omega=9.9998 \times10^{-3}(\hat{\omega}^2=0.99996),\mu(=0.01\ll l),l=1,\nu=1.58,k=6.3\times 10^{-5}$ represents a tiny well ($\bullet$) located at $(r_0=20000,0.99995)$. Here, we include $\hat{\omega}^2=0.999982(2.36)$ and $r\in[100,10^6]$. Other energies are given by the energy quantization Eq.(\ref{omega-R}). (Right) Negative potential $10^{18}V_{\rm h}(r/a_0)$ with $l=1$ and $a_0=1$ always denotes a hydrogenic bound state because of $E<0$. Here, we have $r\in[0.1,100]$. A local minimum ($\bullet$) is located at $(r_0=2,-1.09)$. All negative  energies $E$ are given by the quantization Eq.(\ref{quant-C}).  }
\end{figure*}
\begin{figure*}[t!]
   \centering
  \includegraphics{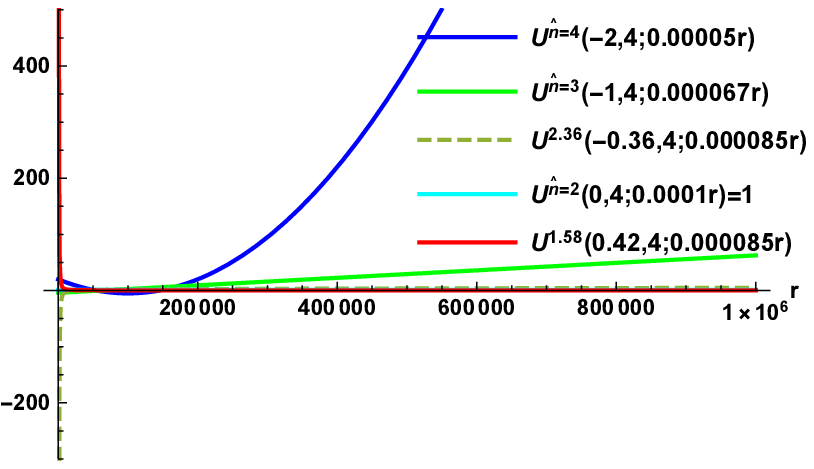}
  \hfill%
  \includegraphics{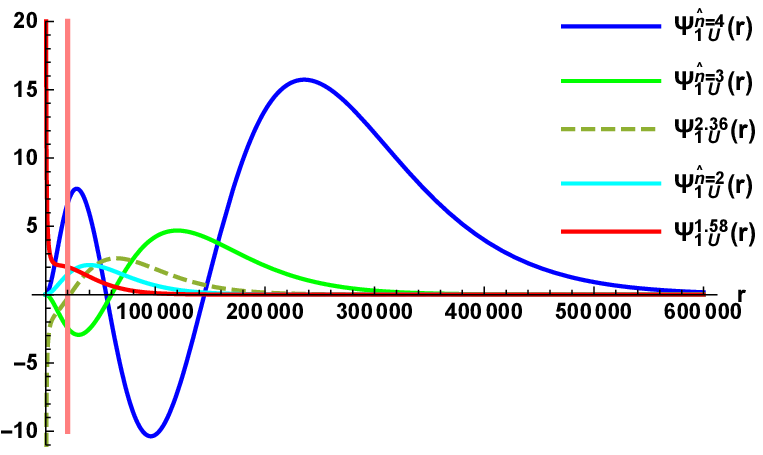}
\caption{(Left) Confluent hypergeometric functions $U^{\hat{n}=4}(-2,4;0.00005 r)$, $U^{\hat{n}=3}(-1,4;0.000067 r)$ and $U^{\hat{n}=2}(0,4;0.0001 r)=1$ in the  region of $r\in[100,10^6]$ for (Left) Fig. 3.  $U^{2.36}(-0.36,4;0.000085 r)$ is a rapidly increasing function, while $U^{1.58}(0.42,4;0.00013 r)$ is a rapidly decreasing function.
  (Right) Hydrogen-like bound states [$\Psi^{\hat{n}=4}_{1U}(r)$,$\Psi^{\hat{n}=3}_{1U}(r)$],  boundary bound state [$\Psi^{\hat{n}=2}_{1U}(r)$], one bound and other with increasing [$\Psi^{2.36}_{1U}(r)$], and a purely  bound state of $\Psi^{1.58}_{1U}(r)$ as function of $r\in[20000$(pink line), $6\times 10^5]$. $\Psi_{1U}^{2.36}(r)$ and $\Psi^{1.58}_{1U}$ are unbound wave functions, because they blow up as $r$ approaches 100 (left boundary).}
  \end{figure*}

Let us choose  the energy quantization  Eq.(\ref{omega-R}) with $\nu=4$. In this case,   one finds the same  potential [(Left)Fig. 3] which includes still a tiny well ($\bullet$).  Here, we choose $\omega=9.99997\times 10^{-3}(\hat{\omega}^2=0.999994)$ which corresponds to the $\hat{n}=4$ quantized state.
The confluent  hypergeometric function  $U(-2,4;0.00005r)$ represents an increasing function [(Left) Fig. 4], while its wave function $\Psi_{1U}^{\hat{n}=4}(r)=2.5\times 10^{-9} e^{-0.000025r} r^2U(-2,4;0.00005r)$ denotes the $\hat{n}=4$ hydrogen-like  bound state [(Right) Fig. 4].
For a complex $\omega=9.99997\times 10^{-3}+10^{-19}i~(|\omega_{\rm I}|\ll \omega_{\rm R},~\omega_{\rm R}<m\Omega_H)$ with the same other parameters as in (Right) Fig. 3, we obtain  its far-region solution
 \begin{eqnarray}
 \Psi_{1U}^{c,\hat{n}=4}(r)&=& (2.5\times 10^{-9} - 8.0 \times 10^{-21} i) e^{(-0.000025 + 4.0\times 10^{-17} i)r}r^{2} \nonumber \\
 &\times& U\Big(\tilde{p}=-2 - 6.4\times 10^{-12}i,4;(0.00005 -8.0\times 10^{-17} i)r\Big), \label{c-wavef}
 \end{eqnarray}
 whose real part Re[$\Psi^{c,\hat{n}=4}_{1U}$] takes the same form of $\Psi_{1U}^{\hat{n}=4}(r)$ as in (Right) Fig. 4.
 Here, one may  rewrite $\tilde{p}$ in $U(\tilde{p},s;cr)$ as $\tilde{p}=-n-\delta\nu$ with $n=-2$ and  $\delta\nu=6.4\times 10^{-12}i$.
 A small complex number $\delta\nu$ represents a deviation from the $\hat{n}=4$ hydrogen-like  energy level. $\omega_{\rm I}=10^{-19}$ may denote a slowly growing superradiant instability,  which states that the scalar grows exponentially in time as $e^{\omega_{\rm I} t}$. \\
We consider   the same  potential [(Left) Fig. 3]  with $\nu=2,3$, corresponding to the $\hat{n}=2,3$  quantized states.
The confluent  hypergeometric function   $U^{\hat{n}=3}(-1,4;0.000067r)$ shows an increasing function, whereas  $U^{\hat{n}=2}(0,4;0.0001r)=1$ represents a constant function [(Left) Fig. 4].  A scalar wave function $\Psi_{1U}^{\hat{n}=3}(r)$ indicates a hydrogen-like bound state, while $\Psi_{1U}^{\hat{n}=2}(r)$ denotes the boundary bound state [(Right) Fig. 4].

Imposing  the condition for  superradiant instability ($\tilde{p}<0$) and stability ($\tilde{p}>0$) for a given potential with a tiny well [(Left)Fig. 3]~\cite{Myung:2022biw},
three [$\Psi^{\hat{n}=4}_{1U}(r)$, $\Psi^{\hat{n}=3}_{1U}(r)$, and $\Psi^{2.36}_{1U}(r)$] show superradiant instability, while $\Psi^{1.58}_{1U}(r)$ indicates  a purely bound state (superradiant stability).
Furthermore, a case of $\Psi^{\hat{n}=2}_{1U}(r)$ represents a boundary between superradiant instability and stability [boundary bound state with $U(0,4;0.0001r)$]. This might be so because
the potential $V_{\rm K}(r)$ in Eq.(\ref{c-pot1}) with a tiny well is an energy ($\omega^2$)-dependent potential.

\begin{figure*}[t!]
   \centering
  \includegraphics{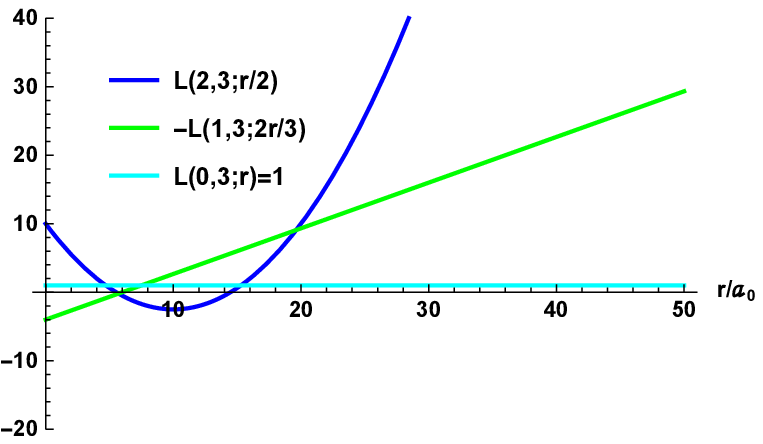}
  \hfill%
  \includegraphics{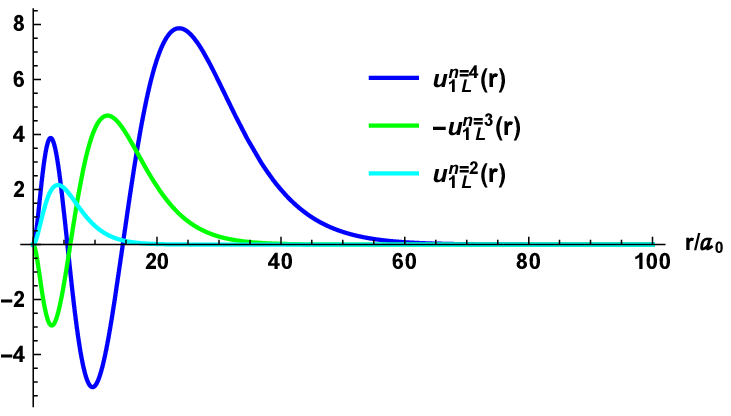}
\caption{(Left) Laguerre polynomials $L(2,3;r/2)$, $-L(1,3;2r/3)=U(-1,4;2r/3)$ and $L(0,3;r)=1$ in the  region of $r\in[0,50]$ for (Right) Fig. 4.
(Right) Three hydrogenic  bound states of  $u^{n=2}_{1L}(r)$, $-u^{n=3}_{1L}(r)$, and $u^{n=4}_{1L}(r)$ for $r\in[0,100]$ with $l=1(p),\rho=2r/(a_0n)$. Number of nodes are 0, 1, 2, respectively. }
\end{figure*}

On the other hand, even though a shape of $V_{\rm h}(r)$ [(Right) Fig. 3] is similar to $V_{\rm far}(r)/\mu^2$, it is a  negative potential.
In the bound states of  $V_{\rm h}(r_0)<E<0$, the electron is inside the potential well and its  energy  is quantized as well as $E<0$.
For $n=2,3,4,l=1,\rho=2r/(a_0n)$, three proper solutions to Eq.(\ref{kumm2}) are given by $z_L=L(0,3;\rho)=1,~-L(1,3;\rho)=-4+\rho,~L(2,3;\rho)=(\rho^2-10\rho +20)/2$ [(Left) Fig. 6].  (Right) Fig. 6 indicates that three wave functions [$u^{n}_{1L}(r),~r\in[0,100]]$  represent the $n=2,3,4$ hydrogenic  bound states in the whole region. Here $u^{n=2}_{1L}(r),~-u^{n=3}_{1L}(r),~u^{n=4}_{1L}(r)$ take  the same forms as for $\Psi^{\hat{n}=2,3,4}_{1U}(r)$ in (Right) Fig. 4 but the latter is defined in the region of $r\in[20000$(pink line), $6\times 10^5]$ far from the outer horizon. Lastly, we point out that $u^{n=1,0,-1,\cdots}_{1L}(r)$ are not allowed for the hydrogenic wave functions because because $u^{n=2}_{1L}(r)$ represents the bottom wave function.

\subsection{Potential without any wells}
We choose a potential with $\mu(=4.14)>l(=4)$ [(Left) Fig. 6] which represents an increasing potential without any positive wells because its local minimum ($\bullet$) is located inside the outer horizon $[r_0(=1.16)<r_+(=1.8$)].
Considering three energies of $\hat{\omega}^2=0.522(\hat{n}=6),~0.407(5.39),~0.311(\hat{n}=5)$, their confluent hypergeometric functions are given by $U(-1,10;5.74r),~U(-0.39,10;6.39r)$, and $U(0,10;6.89r)=1$.
All of their wave functions [$\Psi_{4U}^{\hat{n}=6}(r),\Psi^{5.39}_{4U}(r),\Psi_{4U}^{\hat{n}=5}(r)$] indicate  purely bound states without node [(Right) Fig. 6].
\begin{figure*}[t!]
   \centering
  \includegraphics{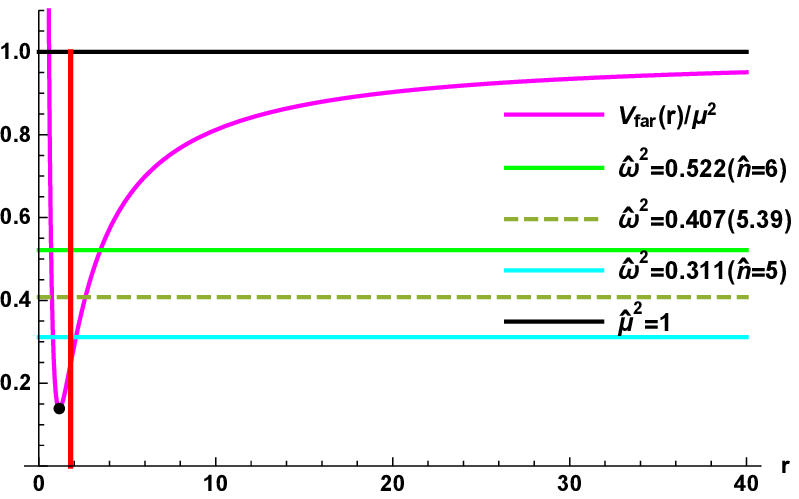}
  \hfill%
  \includegraphics{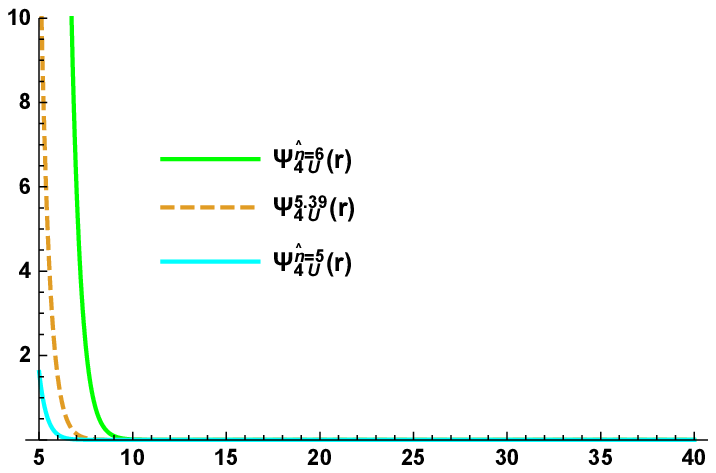}
\caption{(Left) Potential $V_{\rm far}(r)/\mu^2$ with $M=1,\omega=2.65(\hat{\omega}^2=0.407),\mu(=4.15>l),l=4,\nu=5.39,k=3.2$ represents an increasing function of $r\in[2,40]$.
Here, a local minimum ($\bullet$) is located at $(r_0=1.16,0.14)$  inside the outer horizon $r=r_+=1.8$ (red line). Other energies $\hat{\omega}^2$ are given by the energy quantization Eq.(\ref{omega-R}).
  (Right) Purely bound states of $\Psi_{4U}^{\hat{n}=6}(r),\Psi^{5.39}_{4U}(r),\Psi_{4U}^{\hat{n}=5}(r)$ are plotted as functions of $r\in[5,40]$ without node.  }
\end{figure*}
For a complex $\omega=2.997 + 10^{-6} i$,
 we obtain  its far-region solution
 \begin{eqnarray}
 \Psi_{4U}^{c,\hat{n}=6}(r)&=& (6.2\times 10^{3} -0.01 i) e^{(-2.87 + 1.0\times 10^{-6} i)r}r^{5} \nonumber \\
 &\times& U\Big(\tilde{p}=-1 - 2.2\times 10^{-6}i,10;(5.74 -2.0\times 10^{-6} i)r\Big), \label{c-wavef}
 \end{eqnarray}
 whose real part Re[$\Psi^{c,\hat{n}=6}_{4U}$] takes the same form of $\Psi_{4U}^{\hat{n}=6}(r)$ as in (Right) Fig. 6.
Here, we rewrite $\tilde{p}$ in $U(\tilde{p},s;cr)$ as $\tilde{p}=-n-\delta\nu$ with $n=-1$ and  $\delta\nu=2.2\times 10^{-6}i$.
 A small complex number $\delta\nu$ represents a deviation from the $\hat{n}=6$ purely bound level.

\section{Summary and Discussions}

\begin{table*}[h]
\resizebox{14cm}{!}
{\begin{tabular}{|c|c|c|c|c|c|c|c|c|}
\hline
Case & $\tilde{p}=l+1-\nu(l)$ &$\nu$ &$\mu^2$&$\omega^2$&$k$ & well& $\omega_{\rm I}>0$&wave function(figure) \\ \hline
F1&$-1(13)$  &$\hat{n}= 15$ & 16  &14.86 &1.067 & trapping well&yes&Peak,HBS(F2R)\\ \hline
F1&$-0.75(13)$  & 14.75 & 16  &14.82 &1.085 & trapping well&no&Peak(F2R)\\ \hline
F1&$0(13)$  & $\hat{n}= 14$ & 16  &14.69 &1.14 & trapping well&yes&BBS(F2R)\\ \hline
F1&$1(13)$  & $\hat{n}= 13$ & 16  &14.49 &1.23 & trapping well&no&PBS(F2R)\\ \hline
F3L&$-2(1)$& $\hat{n}= 4$& $10^{-4}$& $9.99994\times 10^{-5}$& $2.5\times 10^{-5}$ &tiny well&yes&HBS(F4R)\\ \hline
F3L&$-0.36(1)$&2.36 & $10^{-4}$& $9.99991\times 10^{-5}$& $4.2\times 10^{-5}$ &tiny well&no& UBS(F4R)\\ \hline
F3L&$-1(1)$& $\hat{n}= 3$& $10^{-4}$& $9.99989\times 10^{-5}$& $3.3\times 10^{-5}$ &tiny well&yes&HBS(F4R)\\ \hline
F3L&$0(1)$& $\hat{n}= 2$& $10^{-4}$& $9.99975\times 10^{-5}$& $5\times 10^{-5}$ &tiny well&yes&BBS(F4R)\\ \hline
F3L&0.42(1)&1.58&$10^{-4}$&$9.9996\times 10^{-5}$&$6.3\times 10^{-5}$ &tiny well&no&PBS(F4R)\\ \hline
F3R&$0,-1,-2(1)$&$n=2,3,4$&$\cdot$ &$E=-0.55,-0.24,-0.14$&$1/(a_0n)$ &negative well & $\cdot$& HS(F5R) \\ \hline
F6L&$-1,-0.39,0(4)$&$6,5.39,5$&17.2&8.98,7.02,5.36& 2.87,3.2,3.45& no well &yes,no,yes&PBS(F6R)\\ \hline
\end{tabular}}
\caption{Summary of all quasibound states.
F1, F3L, and  F3R refer Fig. 1, (Left) Fig. 3, and (Right) Fig. 3, respectively.
Peak denotes  resonance and  HBS  represents hydrogen-like bound state which allows to possess $\omega_{\rm I}>0$ based on the energy quantization Eq.(\ref{omega-R}). BBS indicates boundary bound state with $U(\tilde{p}=0,s;cr)=1$ and it corresponds to HBS with $l=1$({\it p}) state.  PBS shows purely bound state without nodes.
Four cases of $\tilde{p}(\nu)=-0.75(14.75),1(13),-0.36(2.36),0.42(1.58)$ represent the unbound wave functions on the left boundary [$\tilde{p}=-0.75$ denotes a peak with increasing function.  UBS($\tilde{p}=-0.36$) means one bound and other with increasing wave function. $\tilde{p}=1,0.42$  show PBS without node.].
HS denotes hydrogenic state and the cases without well  imply  PBS.   }
\end{table*}
We have investigated  all quasibound states of a massive scalar around the Kerr black hole thoroughly  by comparing with a hydrogen atom in quantum states.
Even though the equation is the same as that for the hydrogen atom, its  potential is positive  and  its energy is positive and continuous, compared to the negative potential and  negative quantized energy for the hydrogen atom [see Fig. 1, (Left) Fig. 3, and (Right) Fig. 3].
For the Kerr black hole background, we have obtained three types of far-region  potentials with a trapping well ($\mu<l$),   a tiny  well ($\mu\ll l$), and  no well ($\mu>l$).
We wish to point out that the presence of a trapping well (tiny well) dose not mean directly superradiant instability (stability) because the potential $V_{\rm K}(r)$ in Eq.(\ref{c-pot1}) is an energy ($\omega^2$)-dependent potential. It has to be justified by a form of  far-region wave function $\Psi(r)$ and a positive $\omega_{\rm I}$.

We have explored the  connection between far-region potential with energy and  explicit form  of quasibound states.
When imposing on the abs condition of $\omega<\mu$, the wave function in the overlapping region is either bounded or unbounded (there is no the inner boundary condition).
We summarize the results for all quasibound states in Table 1.
The quasibound states of a massive scalar around the Kerr black hole include  a peak (resonance),  hydrogen-like bound state (HBS) which allows to possess a positive $\omega_{\rm I}$ for superradiant instability,
boundary bound state (BBS) with the constant confluent hypergeometric function $U(0,s;cr)=1$, and purely bound state (PBS) without node.
In general, the superradiant instability for potential with a trapping well implies a far-region wave function of a  peak (resonance)~\cite{Zouros:1979iw,Myung:2022biw,Myung:2022yuo}.

In the case of an ultraviolet boson, however, there is no trapping well but a tiny well is located for from the outer horizon.
In this case, the superradiant instability could be achieved if the following four condition are satisfied~\cite{Kodama:2011zc}:\\
i) $\omega_{\rm I}$ is positive. \\
i) The wave function is asymptotically bounded ($\omega_{\rm R}<\mu$) but it is not peaked.\\
  ${}$ Also, it does not blow up on the left boundary in the overlapping region. \\
ii) $\omega$ is nearly real, implying $|\omega_{\rm I}|\ll \omega_{\rm R}$. \\
iii) $\omega$ satisfies the superradiance condition of $\omega_{\rm R} <\omega_c$. \\
Therefore, three hydrogen-like bound states $[\tilde{p}(\nu=\hat{n})=-1(15),-2(4),1(3)]$ satisfying the energy quantization Eq.(\ref{omega-R})  possess $\omega_{\rm I}>0$, implying that the superradiant instability could be achieved for massive scalar and ultralight boson.
Here, $\omega_{\rm I}=10^{-19}$ describes a slowly (exponentially) growing superradiant instability of the ultralight boson ($\Phi\sim e^{\omega_{\rm I} t}$).
It might produce a gravitational atom which consists of the Kerr black hole and  a surrounding boson cloud in quantum states due to superradiant instability.

Introducing  the condition for  superradiant instability ($\tilde{p}<0$) and stability ($\tilde{p}>0$) for two given potentials with well [Fig. 1and (Left)Fig. 3]~\cite{Myung:2022biw}, five cases of  negative $\tilde{p}~(=-1,-0.75,-2,-1,-0.36)$ represent
superradiant instability, while two positive $\tilde{p}~(=1,0.42)$ indicate superradiant stability (purely bound state without node). Importantly, the former includes three cases of HBS with $\omega_{\rm I}>0 ~(\tilde{p}=-1,-2,-1)$.
In addition, two cases of $\tilde{p}=0$ represents the boundary between superradiant instability and stability [boundary bound state with $U(0,s;cr)$]. This might be so because
the potential $V_{\rm K}(r)$ in Eq.(\ref{c-pot1}) is an energy ($\omega^2$)-dependent potential.

Furthermore, for four cases of $\tilde{p}~(=-0.75,1,-0.36,0.42)$, they have the unbound wave functions on the left boundary in the overlapping region and $\omega_{\rm I}>0$ is not allowed.
These cases are allowed because we do not impose the inner boundary condition. In  case of the  hydrogen atom, the inner boundary condition states that the wave function is regular at the origin. The inner boundary condition corresponds to radiation down the black hole when obtaining  hydrogen-like bound states~\cite{Detweiler:1980uk}.
If one requires the inner boundary condition of bound states in the overlapping region, the four cases are eliminated from the spectrum.

Finally, we confirm  three hydrogenic  bound states shown in (Right) Fig. 5 for the hydrogen  potential and negative quantized energy in (Right) Fig. 3.
If there are no wells outside the outer horizon, their far-region wave functions all represent purely  bound states without node.

 \vspace{0.5cm}

{\bf Acknowledgments}
 \vspace{0.5cm}

This work was supported by a grant from Inje University for the Research in 2021 (20210040).

\end{document}